# On signatures of sonic wavepackets in time-resolved X-ray diffractometry of metal single crystals absorbing pulse from ultrafast laser

**Oleg Korovyanko[1] and Oleksandra Korovyanko[2]**

[1]Department of Solid State Physics, [2]Department of Chemistry,

Chernivtsi National University, Chernivtsi, 58000 Ukraine

Abstract

Copper (Cu), gold (Au) (111) crystals were illuminated with 120 fs pulses and probed by 600 fs X-ray pulses. Rocking curves were measured versus 267 nm (UV) pump- 0.154 nm ($CuK_\alpha$) probe delay time. Curve width broadening, peak diffracted intensity and angular shift were recorded for the range of UV excitation intensities of 2-8 mJ/cm$^2$ . Observed oscillations in time delay dependences of shift, width and of peak intensity of rocking curves are correlated, as described by acoustic pulse bouncing between crystal surfaces. In (111) Au, acoustic pulse originates from pump absorption layer, and peak intensity drops by a factor of two (at delay time at which sonic wave reaches Au-substrate boundary) of its initial value prior to absorption of laser pulse. In Cu, correlated rocking curves dynamics is also observed, however buildup of rocking curve broadening is faster than one can expect from sonic wavepacket originating in pump absorption layer. This effect is attributed to acoustic phonon generation outside of pump absorption layer, it is much stronger in Cu than in Au. Phonon generation in Cu is distributed throughout the whole bulk of crystal due to delocalization of mobile hot electrons diffusing from pump absorption layer.

Sound wave generated by absorption of light with optical penetration depth $\delta$, typically is expected to have wavelength $\lambda_s \sim \delta$ [1]. Typical X-ray extinction length $\tau_{ex}$ in noble metals, CuK$_\alpha$ (111) X-ray Bragg reflection, is of order of several $\mu$m [2]. Pump pulse penetration depth is an inverse of absorption coefficient $\alpha$ at 267 nm, $\delta\,(\alpha^{-1})$ is around 13 nm. This sets the ratio of sound wavelength to extinction length $\lambda_s/\tau_{ex} < 1$; this ratio determines the nature of x-ray diffraction on acoustic super-lattice. The UV pulse launches acoustic wave-packet with typical period of sound wave in 10-100 ps range [1]. Studies of generation and propagation of such waves are performed using subpicosecond X-ray probes [3-7].

Time-resolved X-ray diffraction (XRD) was used in many structural studies of metals and semiconductors. In metal single crystals femtosecond laser pulses interact with electrons, that absorb photons. Electron gas reaches partial-equilibrium state via electron-electron interaction on time scale of 120 fs, typical temperature values reach $10^4$ K [1-7]. Electron gas is initially out of equilibrium with lattice. Electron gas relaxes its energy to the phonon modes of lattice on time scale of 10-100ps [2]. Lattice expansion and contraction, strain, sonic and blast waves were revealed in previous studies of gold and silver crystals [3-7]. Mean free path of electrons in Cu is higher than that in gold and silver, likely due to higher electron mobility and smaller electron-phonon coupling constant of $g \approx 20\text{–}60 \times 10^{15}$ W$\cdot$m$^{-3}\cdot$K$^{-1}$J [9].

At sufficient excitation intensity melting of crystals was also observed, followed by formation of mosaic polycrystalline structure with subsequent re-crystallization [5]. Copper has higher conductivity than gold and silver, and electron gas relaxation is a subject of on-going studies [9].

In this paper we present results of femtosecond transient Bragg diffraction from single copper (111) and gold (111) crystals. We detect the XRD rocking curves including peak shift, broadening, and peak diffraction intensity change following sub-picosecond pulsed UV irradiation. The study reveals evolution of coherent acoustic pulse. We observe correlated dumped oscillations in time dependences of Bragg peak intensity and of shift and width broadening of rocking curves. The integrated diffraction intensity does not change as predicted [2] for sound standing waves below melting threshold in metal single crystals.

Cu (111) 90 nm thick crystal samples and 150 nm Au (111) crystals were grown on mica substrate. Bragg reflection maximum was found at 20.2 degrees with respect to Cu (111) input surface, with half-maxima of angular dependence I ($\Theta$) at 19.6 and 20.8 deg. We have found that Cu(111) planes are tilted by about 1.4 deg in respect to the crystal surface. In Au (111) crystal Bragg reflection maximum was at 19.3 degrees, with half-maxima of angular dependence

I ($\Theta$) at 18.9 and 19.6 deg. The crystal was illuminated with 120fs laser pulses from compressed output of multipass Titanium-sapphire amplifier [8] frequency-tripled to wavelength of 267 nm (UV), maximum excitation pulse energy was ~10 mJ/cm$^2$. Nearly 25% was reflected from sample surface, absorbed portion of the pulse energy was 2.1 to 7.5 mJ/cm$^2$. CuK$_\alpha$ X-ray pulses were generated from copper wire target hit by femtosecond pulse, the same way as in ref. [4-6]. The total number of photons per pulse was estimated to be ~10$^7$ per full 4$\pi$ stereorad body angle. UV pump beam was focused to an elliptical spot of 2.2x1.1 mm on the sample surface. 300 $\mu$m vertical slit was placed at 5 cm from the wire, in order to limit x-ray beam. The angular width of X-ray beam before incidence onto (111) crystal surface was ~30 mrad.

The Bragg angle at the Cu (111) plane for CuK$_{\alpha 1}$ reflection is 21.5 deg. The rocking curves were recorded with a CCD for 15 min at each delay time between UV and X-ray pulses set by mechanical translation of retro-reflector. We used three 1.8mm CCD sections 30 pixels each along a line normal to x-ray plane of incidence, one reflected from heated area of sample in the middle and two from un-heated sections below and above. Each point of rocking curves contained total of readings of 30 CCD pixels. Rocking curves were fit to Gaussian curves, fit parameters of rocking curves from illuminated areas were normalized using two rocking curves from adjacent areas not illuminated by UV pulse. Zero time delay $\tau$=0 (X-ray and UV pulses arrive to sample at the same time) was determined by repeating pump-probe experiment described in Ref. 6 using single gold (111) crystal as a sample. We measured rocking curves at delay times $\tau$ from 0 to 120 ps.

In present study femtosecond UV pump pulse illuminates copper 111 single crystal. Alternatively, same pump and probe pulses were applied to gold (111) crystal with $\delta$ =12 nm.

Cu samples were initially tested with atomic force microscope (Fig.1). The scan started from the uncoated mica substrate and continued over the step on the border line of Cu crystal coated on top of mica, see image in Fig. 1B. Fig.1 A shows AFM tip position versus spatial coordinate X across Cu (111) single crystal border line at X$_b$=13 $\mu$m along the line shown in white in Fig. 1b. Left part shows scan on mica substrate (X<13 $\mu$m), whereas right side (X<13 $\mu$m) shows profile of copper surface. The profile can be roughly described by a step function with a step size (crystal thickness) of ~90 nm.

Two rocking curves of (111) CuK$\alpha$ reflection from Cu sample are shown in Fig 2, they were recorded from the same area of Cu crystal The spatial shift of center of Gaussian fit is marked Shift, it is~3 mrad along $\Delta\Theta$ axis. If normalized, the curves were almost overlapped on higher-angle side. The width of right curve at half maximum (FWHM$_{no\ UV}$) is 12 mrad. The left curve (B) is ~50% broader (FWHM$_{UV}$ ~18mrad). Estimate of average rocking curve expansion gives FWHM$_{UV}$ - FWHM$_{no\ UV}$ =6 mrad. It is referred to as broadening, or rocking curve broadening, in this paper. Dynamics of broadening w(t) in Cu crystal is shown in Fig.3a. Offset of w(t) initial increase (see also Fig.3a inset) occurs at delay time $t_{off}$=-6ps. From the same figure, w(t) reached maximum at t=6ps. Taking offset into account, the pump-probe delay $\tau$ is t - $t_{off}$= 12 ps. w(t) in Fig 3a is a sinusoid-like curve with damped amplitude. Other maxima are observed at 46 and 87 ps, with corresponding minima between maxima at 24 and 66 ps.

Rocking curve at the left side of Fig.2 was recorded by CCD at $\tau$=12 ps following pump pulse hits the sample. Position of CCD pixels are transformed to angle using sample-to CCD distance of 15 cm. Maximum (peak) intensity decreases by ~36% whereas the angular half-width increases by 50 % (Fig.2). The curves are gaussian fits of rocking curves at $\tau$=12 ps and at at $\tau$=-40 ps.

Direct comparison of peak intensities of two curves in Fig.2 gives 33000 versus 22000 arb. units.

In Fig.2 rocking curve shifts to smaller Bragg angle, upon pump absorption. Taking into account Bragg condition 2**d** sin $\Theta$=$\lambda$ for X-ray, we can differentiate both sides to get $\Delta\Theta$ / tan $\Theta$=$-\Delta$**d/d**

i. e. we expect smaller $\Theta$ at larger lattice interplane spacing **d**. Taking Shift $\Delta\Theta_{max}$ =$-$3 mrad from Fig.2 (marked Shift), we expect lattice expansion $\Delta$d to reach 1.6x10$^{-2}$ Angstrom. Taking into account linear thermal expansion coefficient of Cu of 1.66x10$^{-5}$ K$^{-1}$ we estimate the lattice temperature increase $\Delta$T to be of 464 K.

Around 50% broadening of rocking curve with UV excitation in Fig. 2 is due to varying expansion of spacing **d**. $\Delta$d/d is highest in crystal layers with highest thermal and/or elastic expansion. $\Theta$ in these layers shifts to lower angles $\Theta$ +$\Delta\Theta$ , where $\Delta\Theta$ < 0, and $\Delta\Theta$ is widely distributed in acoustic wavepacket with variable $\Delta$**d/d** , $\Delta\Theta_{max}$< $\Delta\Theta$ < 0. Zero $\Delta\Theta$ corresponds to undistorted layers in the crystal. Absorption depth of pump pulse $\delta$ is ~13 nm in Cu, typically is expected to be an order of magnitude lower than X-ray penetration depth $\delta_{X-ray}$, 1/($\alpha_X$ cos$\Theta$)~1$\mu$m, $\alpha_X$ -CuK$\alpha$ absorption coefficient in Cu metal. Therefore, a lot of undistorted crystal layers diffract CuK$\alpha$ X-ray, even with UV pump excitation at $\tau$=2 to 120 ps .

As a result of pump absorption and subsequent lattice expansion with generated sonic wave the broadening w widens at $\tau$>2 ps (Fig.2). By the same token, I($\Delta\Theta$) decreases. Integrating I-V($\Delta\Theta$) after taking background $I_0$ such as in Fig. 2, over $\Delta\Theta$, we obtain variation in integrated intensity (not shown). I did not change within experimental error, it is roughly constant with delay time $\tau$.

If the crystal lattice is distorted with acoustic wave with wave number $k_s$, total reflected intensity becomes [2]

$$I \sim \left|\sum_{n=1}^{\infty} \exp(ikr)\right|^2 \sim \left|\delta_{\Delta k, G} + i\Delta ku \left(\delta_{\Delta k+ks, G} e^{-i\omega t} + \delta_{\Delta k-ks, G} e^{i\omega t}\right)\right|^2 \quad (1)$$

$k$, $\Delta k$ - X-ray wavevector parameters on dispersion curve, $\omega$ - X-ray ($CuK_\alpha$) angular frequency, u- lattice displacement due to sonic wave, $\delta$ - delta-function. If satellites $k \pm k_s$ do not overlap, the cross term vanish, and there is no time dependence.

Fig. 3b shows peak X-ray (111) diffraction intensity I(t). data points are recorded twice at the same delay times t in order to estimate error in experiment. The solid curves in Figs 3, 4(b) and 5 are guides to the eye. Following initial drop of ~50% at t=6ps ($\tau$ = 12 ps), two full I(t) oscillations are observed, with minima at t = 6 and 40 ps, and a maximum at t=24 ps (Fig.3b). Comparing Fig.3a and 3b, we see that w(t) and I(t) show two oscillations with opposite phases. Rocking curve shift (see Shift(t) in Fig.4a) and broadening (Fig.4b) were measured in 150 nm (111) gold. From Fig.4, Shift(t) maxima are at t=55 and 158ps, whereas its minimum is at t=100ps. The positions of w(t) maxima (at 54 and 155 ps) and minimum at 98 ps are in agreement with positions of maxima and minimum in Fig.4a. The $t_{off}$=2ps in Fig.4a,b.

w(t) dynamics in 90 nm (111) Cu (Fig. 5b) nearly reproduces dynamics of Bragg angle Shift (Fig.5a), $|\Delta\Theta|$. At moderate pump intensity of 2.5 mJ/cm$^2$, we see that Shift (t) and w(t) are nearly correlated oscillatory curves, w(t) shows oscillations with more pronounced damping (compare Fig 5, a and b).

The picture interpreting w(t) and Shift (t) in Figs. 2-5 is the following. Many-body interaction of photons with free electrons results in heating of electrons. Hot electrons interact within electron gas reaching equilibrium temperature about 10$^4$K. Electron gas exchanges energy with lattice vibration (phonons). Phonons are affecting rocking curves, coherent "breathing" is the result of superposition of diffraction patterns of phonons with frequencies distributed as an acoustic pulse generated according to boundary conditions.

On time scale of electron-phonon thermalization (1-2 ps) hot electrons appear delocalized far from light absorption length (skin depth) [1,9]. Following photoexcitation, they move in Cu with velocities of order of 10$^7$m/s, mean free path length $L_F$ is 72 nm, with average hot electron penetration depth of order of $L_F$ / 2 =36ps. Large electron temperature gradients are formed at the end of free path, where 10-20 % percent of hot electrons give their energy to coherent phonons. Coherent phonons can reach copper-mica boundary in time $(L-L_F)/v_s$. As a result, in Fig 3 first maximum in width broadening plots w(t) occurs at lower delay time $\tau=54$ ps, and time intervals between w(t) maxima (or I(t) minima) are 40 and 41 ps (Fig. 3a) or 36 ps in Fig. 3b. In gold first w(t) peak occurs at $\tau=55$ ps (Fig.4b), with first Shift(t) maximum at $\tau=54$ ps (Fig.4a). Time interval between w(t) maxima (or Shift(t) maxima) is 103 ps (Fig. 4a) (or 101 ps in Fig. 4b). Those time intervals are in agreement with round trip time of sound waves in gold, $\tau_{extr,i} = 2iL/v_s$, i=1,2,3  2x150nm/3.05km/s=98 ps; 2x90nm/4.7km/s= 38 ps, or in gold.

The oscillatory character of w(t), I(t) and Shift (t) in metals is in good agreement with model of sound wave packet, bouncing between air-metal and metal-substrate boundaries. Time interval between sound reflections in our experiment is in agreement with previous studies [4-7]. However, in (111) Cu first extremum of w(t), I(t) and Shift (t), $\tau_{extr,1}$ occurs at much earlier delay time than that predicted by the same model, at $\tau=12$ps rather than at $\tau_{extr,i} = L/v_s = 19$ ps. For possible reasons, we consider first reflection of expansion wave from metal-mica boundary. w(t), I(t) and Shift(t) go through extremum (Fig. 3-5), when the expansion wavepacket does not have any more (111) planes to expand into, i.e. wavepacket reflects from the substrate boundary. At the time of reflection a portion of sonic pulse reflected at nodes interfere constructively with incident pulse, the remainder pulse has a random phase at reflection. Following reflection at metal-mica boundary w(t) and Shift (t) decrease until another minima corresponding to $\tau$ equal to round trip $2L/v_s$ of sound wavepacket. This pattern reproduces itself at times of subsequent second, third reflections etc.

We will estimate a distance over which expansion wave propagates before $\tau_{extr,1}$ occurs in Cu as 12ps x 4.7 km/s = 56 nm. It is far from the pump absorption layer with depth of 13 nm, at 90nm-13nm=77nm distance from copper-mica boundary. We suggest that in average electron-phonon scattering occurs at depth of 90nm-56nm=34 nm. This number is in agreement with average penetration depth of hot electrons in copper $L_F / 2$ =36ps.

To conclude, we measured time-dependent X-ray rocking curves of Cu (111) and Au (111) single crystals with sub-ps pulsed UV excitation. The model of sonic wave bouncing between air-metal and metal-substrate [3-7] satisfactorily explains time intervals between correlated w(t), I(t) and Shift (t) extrema $\tau_{extr,i}$ with i=1,2,3 etc., both in Cu and Au. Different models were applied to explain early occuring extremum $\tau_{extr,1}$ in copper in contrast to regular $\tau_{extr,1} \sim L/v_s$ in gold. We conjecture that weaker electron-phonon coupling and higher electron mobility in copper results in generation of phonons throughout mean free path length $L_F$ of hot electrons heated by pump laser pulse. We measured transient Bragg diffraction in silver films with $L_F$ limited by crystal grain boundaries [10].

Acknowledgement

The author acknowledges Tetyana Ignatova for AFM measurements, Mariana Borcha for discussions and Peter Rentzepis for kindly sharing his laboratory equipment. The author's work at University of California, Irvine was supported in part by the W. M. Keck Foundation.

References

1. Akhmanov S A, Gusev V E, Laser excitation of ultrashort acoustic pulses: New possibilities in solid-state spectroscopy, diagnostics of fast processes, and nonlinear acoustics, *Sov. Phys. Usp.* **35** (3) 153–191 (1992).

2. Ultrafast Lattice Dynamics, A.M. Lindenberg, in *Nonlinear Optics, Quantum Optics, and Ultrafast Phenomena with X-rays, edited by B. Adams, Springer Science + Business Media, New York* (2003).

3. J. Wark, A.M. Allen, P.C. Ansbro, P. Bucksbaum, Z. Chang, M. DeCamp, R.W. Falcone, P.A. Heimann, S.L. Johnson, J. Kang, H.C. Kapteyn, J. Larsson, R.W. Lee, A.M. Lindenberg, R. Merlin, T. Missalla, G. Nay, H.A. Padmore, D.A. Reis, K. Schejdt, A. Sjoegren, P.C. Sondhauss and M. Wulff, Femtosecond X-Ray Diffraction: Experiments and Limits, *Proceedings of SPIE* 4143, 26 (2001).

4. J. Chen, I. V. Tomov, H. E. Elsayed-Ali, and P. M. Rentzepis, Hot electrons blast wave generated by femtosecond laser pulses on thin Au(111) crystal, monitored by subpicosecond X-ray diffraction Chem. Phys. Lett. 419, 374–378 (2006).

5. Chen, J.; Chen, W.-K.; Tang, J.; Rentzepis, P. M. Time-Resolved Structural Dynamics of Thin Metal Films Heated with Femtosecond Optical Pulses. Proc. Natl. Acad. Sci. U.S.A., 108, 18887 (2011).

6. Ali Oguz Er, Jie Chen, Jau Tang, and Peter M. Rentzepis Coherent acoustic wave oscillations and melting on Ag(111) surface by time resolved x-ray diffraction, Appl. Phys. Lett. 100, 151910 (2012) and references therein.

7. Subpicosecond and Sub-Angstrom Time and Space Studies by Means of Light, X‑ray, and Electron Interaction with Matter, Jie Chen, and Peter M. Rentzepis, J. Phys. Chem. Lett. 5, 225−232 (2014).

8. Positive Light Inc

9. M. Lejman, V. Shalagatskyi, O. Kovalenko, T. Pezeril, V. V. Temnov, and P. Ruello, J. Opt. Soc. Am. B 31, 282 (2014)

10. O.Korovyanko, to be published

Figure Captions

Fig.1. Cu (111) single crystal border - air profile measured with atomic force microscope (AFM).

a) profile across the copper crystal boundary along white line in (b), starting from bare mica substrate (left side) and continuing on crystal surface (right side); b) image from the AFM microscope.

Fig. 2. Rocking curves in Cu(111) prior to (scatter with filled circles, on the right) and 12 ps following photoexcitation (scatter with filled squares, on the left) at UV intensity of 5mJ/cm$^2$. Rocking curve Shift and broadening are defined.

Fig.3. Rocking curve width broadening w(t) (a) and peak intensity I(t) (b) for Cu(111) crystal as a function of delay time, at excitation UV intensity of 5 mJ/cm$^2$, multiple datapoints along y-axis show experimental error. Inset in (a) shows offset of growing w(t) with t-axis step size of 3 ps.

Fig.4. Shift of rocking curve (a) and rocking curve broadening in Au(111) crystal at UV intensity of 2.5 mJ∕cm$^2$, as a function of delay time. Bars in (a) show experimental error

Fig.5. Shift of rocking curve (a) and rocking curve broadening in Cu(111) crystal at intensity of 2.5 mJ∕cm$^2$, as a function of delay time.

Fig.1

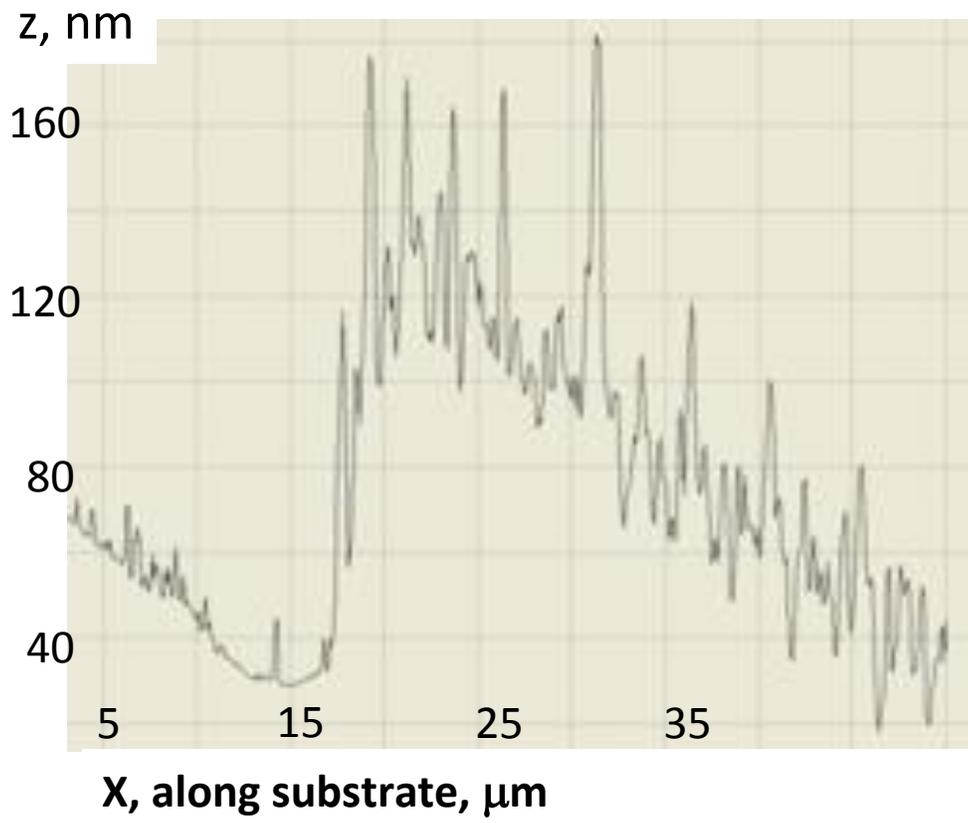

X, along substrate, μm

A

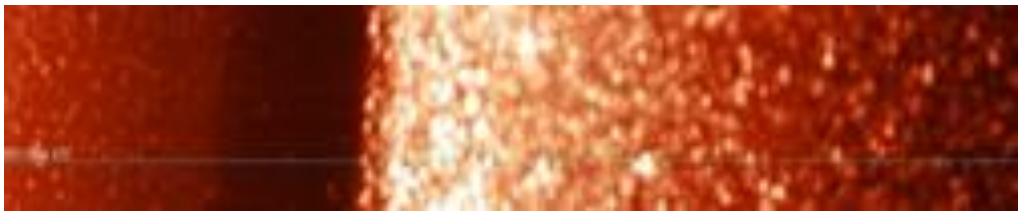

B

Fig.2

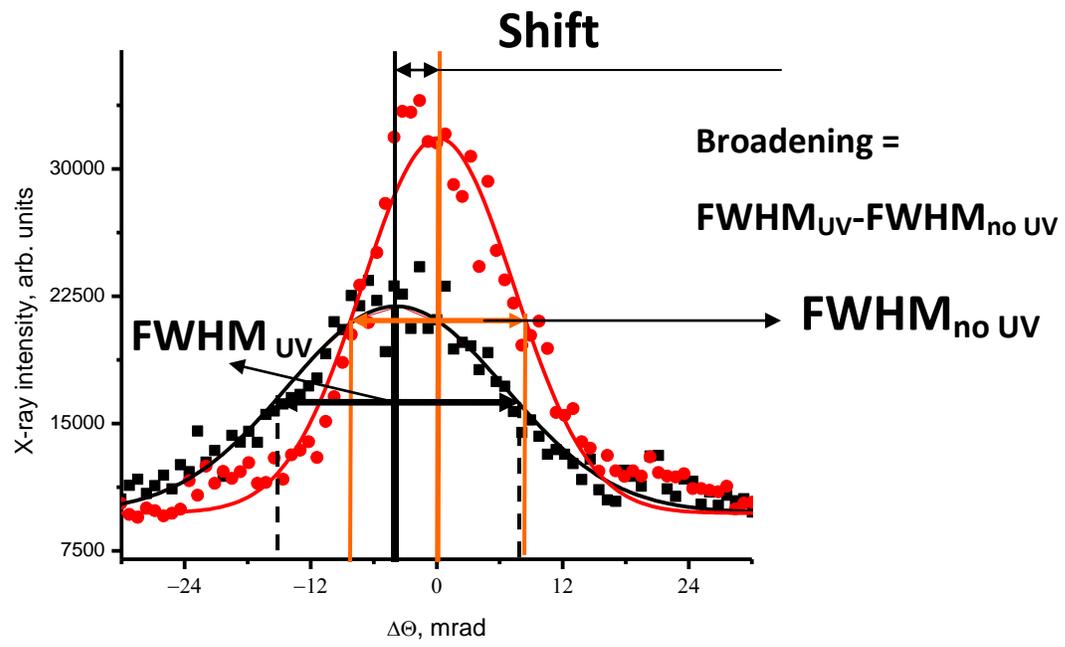

Fig.3

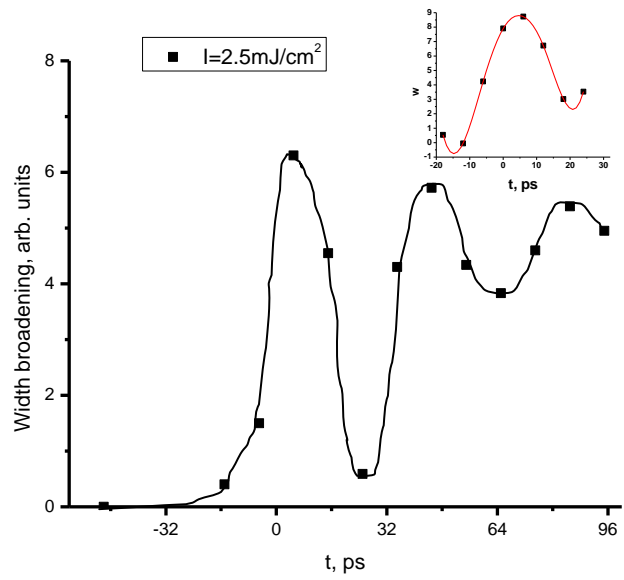

A

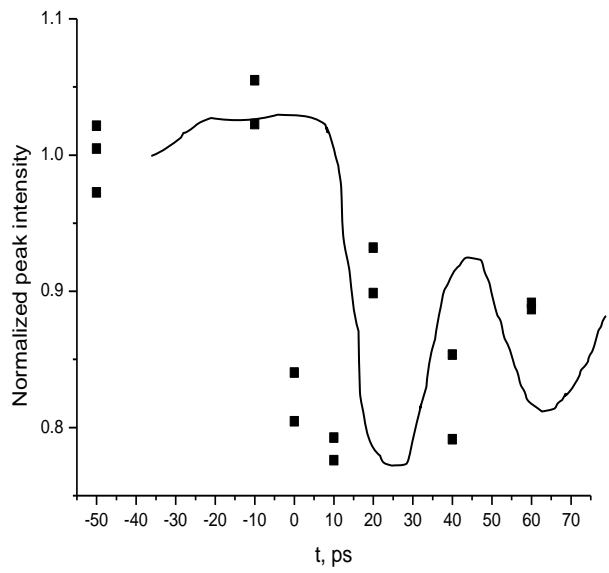

B

Fig.4

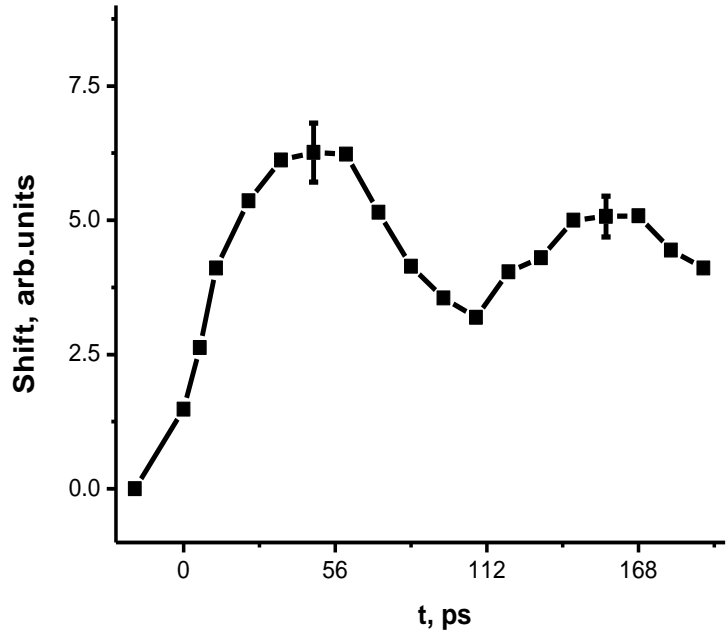

A

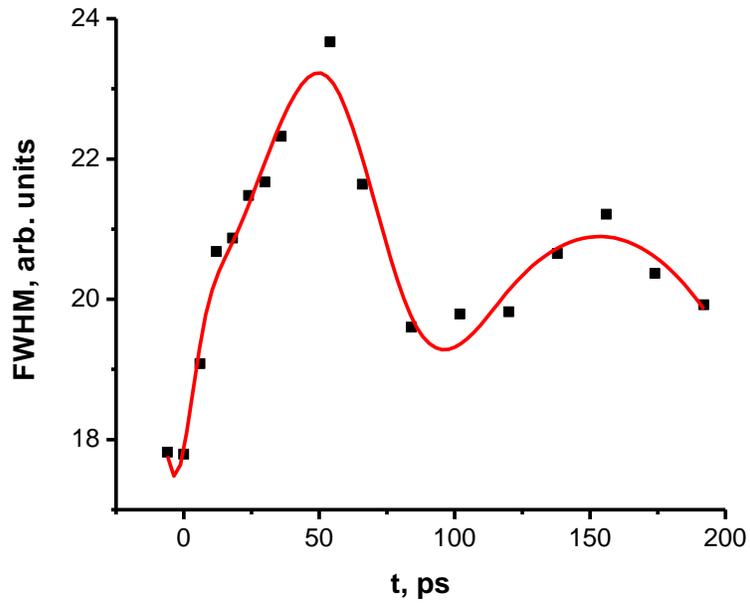

B

Fig.5

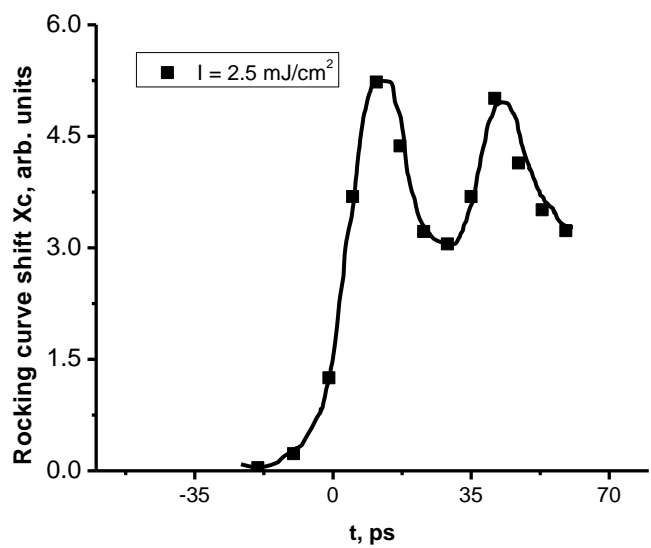

A

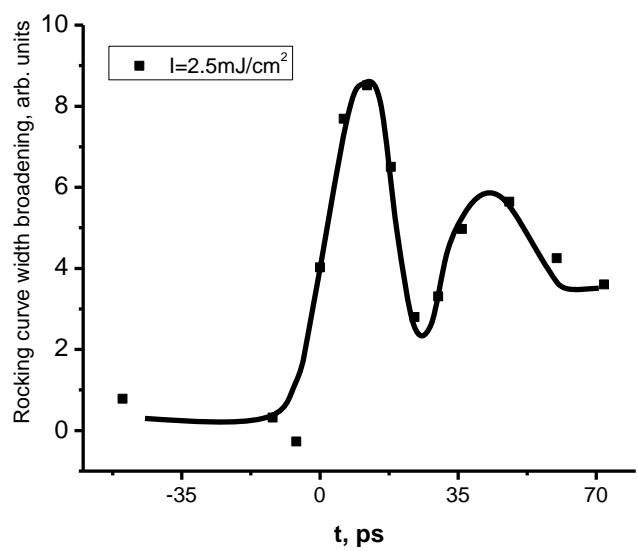

B